# Drift Removal in Plant Electrical Signals via IIR Filtering Using Wavelet Energy

Saptarshi Das[1*], Barry Juans Ajiwibawa[1], Shre Kumar Chatterjee[1], Sanmitra Ghosh[1], Koushik Maharatna[1], Srinandan Dasmahapatra[1], Andrea Vitaletti[2], Elisa Masi[3], and Stefano Mancuso[3]

1) School of Electronics and Computer Science, University of Southampton, Southampton SO17 1BJ, United Kingdom.

2) DIAG, Sapienza University of Rome, via Ariosto 25, 00185 Rome, Italy.

3) Department of Agri-food Production and Environmental Science, University of Florence, Florence, Italy.

Email: sd2a11@ecs.soton.ac.uk, s.das@soton.ac.uk (S. Das*), Phone: +44(0)7448572598

**Abstract**

Plant electrical signals often contains low frequency drifts with or without the application of external stimuli. Quantification of the randomness in plant signals in a stimulus-specific way is hindered because the knowledge of vital frequency information in the actual biological response is not known yet. Here we design an optimum Infinite Impulse Response (IIR) filter which removes the low frequency drifts and preserves the frequency spectrum corresponding to the random component of the unstimulated plant signals by bringing the bias due to unknown artifacts and drifts to a minimum. We use energy criteria of wavelet packet transform (WPT) for optimization based tuning of the IIR filter parameters. Such an optimum filter enforces that the energy distribution of the pre-stimulus parts in different experiments are almost overlapped but under different stimuli the distributions of the energy get changed. The reported research may popularize plant signal processing, as a separate field, besides other conventional bioelectrical signal processing paradigms.

## 1. Introduction

Processing and analysis of biological signals like Electrocardiogram (ECG), Electroencephalogram (EEG), and Electromyogram (EMG) commonly precedes by a pre-processing stage for removing low frequency artifacts and high frequency noise [1]. Selection of appropriate pre-processing technique, in these cases, is driven by extensive understanding of the nature of the signals, resulted from decade-long explorations. In the recent years, interests in understanding the behavior of plants under different stimuli and using that behavior in analyzing different environmental condition [2] and its application in robotics [3] have been growing. The fundamental requirement in such cases is to analyze the electrophysiological signals, recorded from plants under different stimuli. However, unlike the typical biological signals, mentioned earlier, there exists no framework for pre-processing of plant electrical signals, mainly due to fact that there is no clear understanding of artifacts and noises in plant signals [4]. As shown in Figure 1, from our experiments (described in section 2), we found that there exists an initial bias in the response, even before a stimulus is applied. Such bias (for convenience termed as *drift*) needs to be removed for reliable analysis of the post-stimulus part of the plant signal. The drift, possibly containing the effect of leaf movement, several other unknown environmental effects (like humidity, temperature), biological condition of the plant, difference due to the recording instrumentation like channel gains etc., may bias the statistical characteristics of the pre-stimulus part of the plant electrical signals. Also, in order to avoid removal of some significant frequency components containing vital information of the actual biological response in the recorded raw plant signal (as a side effect of filtering), there is a need to investigate an optimal bandwidth of its frequency spectrum. The plant signals are generally weak in nature [5], and are therefore prone to get overwhelmed by noise, artifacts and drifts, which motivates the design of a robust filtering technique, to get rid of these external disturbances. This will ensure that the pre-stimulated signals, recorded under different experimental conditions lie in a common platform, for the study of the stimulated response in plants.

### 1.1. Background and motivation

Standard analysis procedure of other bio-signals e.g. EEG analysis shows that the vital cognition related biological responses lie in the high frequency regimes than the lower frequencies. This motivates us to search for an optimum frequency band within the whole spectrum which does not discriminate signals, based on plant specific and experimental condition specific characteristics. The present filter design for plant electrical signal processing is attempted with the motivation to have minimum discrimination in the unstimulated signals – even if they might come from different plants, channels etc. We found that the strong low frequency components which often contains the characteristics of the drift and artifact (in time domain) are not consistent in different channels and different plants. The presence of these low frequency drifts discriminate the background (pre-stimulus) signal for different plants and channels as shown in Figure 1. This makes the task of discriminating the stimulated plant responses quite difficult because the true response often gets buried under strong low frequency drifts or trends. These inconsistent low frequency drifts (trends) must be removed so that the most consistent statistical characteristics in the optimum frequency bandwidth might be focused on, for the discrimination of the externally applied stimulus. Therefore, the purpose of the filter design here is to detrend the raw plant signals as a pre-processing step for extracting other informative statistical features from the detrended signal, containing high frequency information which are believed to be worth preserving, similar to the case of other biological signals [1].

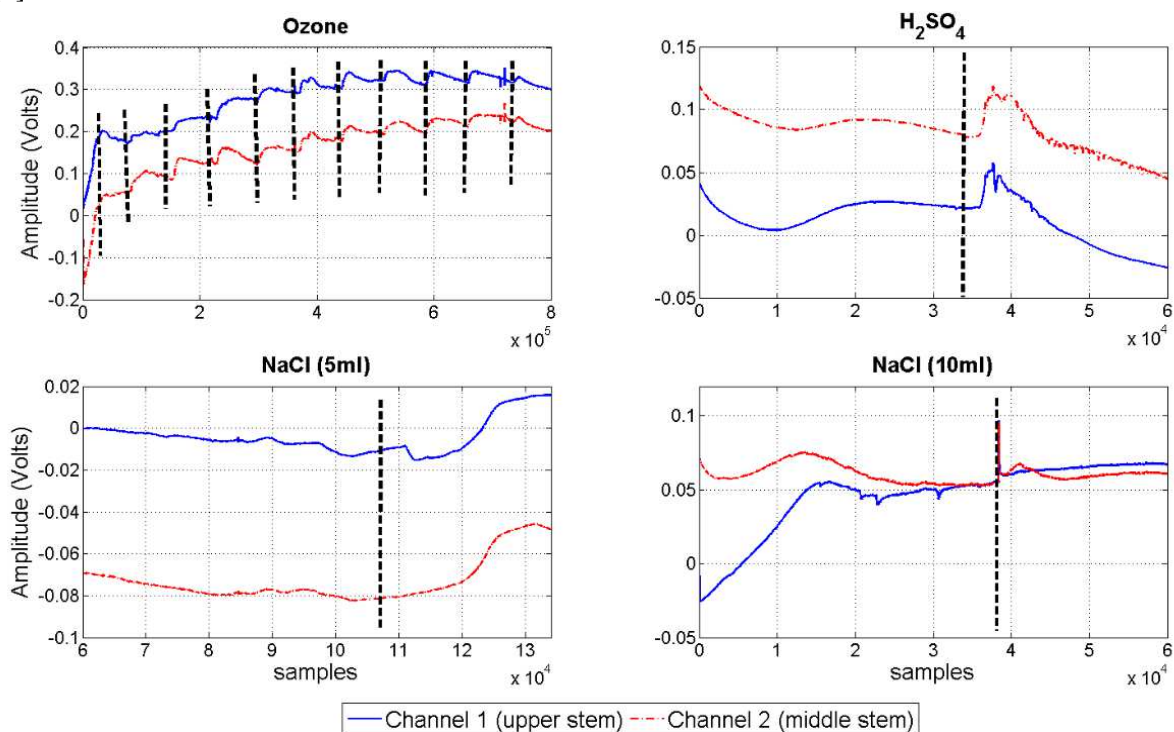

Figure 1: Plant electrical response due to four different external stimuli. Dotted vertical lines indicate the time when stimulus was applied.

In this paper, we explore the optimal settings of four classes of digital IIR filters for plant electrical signal processing and have reported a comparison amongst them. Our approach is based on the concept that some statistical characteristics of the unstimulated plant signal (after filtering) should be similar e.g. zero mean, uniform variance and energy etc., as commonly employed for processing of other bio-potentials as well [1]. It will help in modelling the random component of the plant signal or the underlying data generating mechanism as an ergodic stochastic process [6]. Ideally, the pre-stimulus part of different plant electrical signals or different ensembles in similar experimental conditions, should form clusters in a statistically meaningful feature space, with their centroids lying as close as possible. Therefore, the IIR filter should be designed in such a way that it makes the pre-stimulus part of the signal overlapped and form a common reference for different plants and recording electrodes. By doing so, any change in the electrical response due to the application of external stimulus can be analyzed as the deviation from this reference or the filtered background signal. Also, since the plant signals show nonlinear input-output relationship [7] and strong non-stationary behavior [8-9], use of wavelets becomes a natural choice, for estimating its frequency domain response. The performance of the frequency domain design of IIR filters are judged by an objective function which uses the non-stationary time-frequency domain decomposition using wavelets. Although it might seem redundant but the present use of wavelets are not as filters but to provide a selection criteria for the IIR filters, having much less number of tuning parameters compared to its counterpart – Finite Impulse Response (FIR) filters.

The objective function for the optimization based tuning of IIR filters uses the energy contents of different nodes in *wavelet packet decomposition* for both the pre- and post-stimulus parts of the plant signal, acquired under similar laboratory settings. We tune the optimal IIR filter parameters such that it produces almost overlapping clusters of the distribution of energy along different wavelet basis for the pre-stimulus signals, but non-overlapping clusters for the post-stimulus signals. We also explore the effect of varying window length (number of data samples) while calculating the spectral energy to get an effective filtering or detrending of the raw plant signals. Finally, we have also explored the variability of the clusters in the wavelet energy domain using different wavelet basis functions.

### *1.2. Previous approaches of plant electrical signal processing*

There have been different opinions amongst researchers regarding whether the essential electro-physiological information of plants lie in the low or high frequency region because in plants many mechanisms are slower than in animals – electrical signals are just one example. In fact, the waveforms of plant signals show relatively long durations and signals propagate more slowly with respect to the animal's ones as shown by Pickard [10], Fromm and Lautner [11]. Accordingly, the frequencies of plant signals are expected to be smaller, as reported in Masi *et al.* [12, 13]. However, for using plant signals to discriminate externally applied chemical stimuli, the optimum bandwidth should be determined using a cost function optimization based approach. The cost function should be designed in such a way that it ensures minimum discrimination of the pre-stimulus signal (using some features – wavelet decomposition energy here), irrespective of the plant species, experimental condition and electrode positioning.

Previous exploration on plant electrical signal processing has been using the cross-correlogram analysis [13] during increased gravity, power spectrum analysis in different duration of osmotic stress [14], time, frequency and time-frequency domain analysis under controllable light with different intensity [15] etc. Previous research divides the plant signal spectrum in arbitrary frequency bands similar to the EEG's for studying power of the bands [14]. Our approach helps in finding the common part of the whole frequency spectrum that does not discriminate between different experimental conditions before any stimulus was applied. Wavelets have also been previously used in Tian *et al.* [15] to obtain detailed and approximate coefficients representing different frequency bands of the signal but not intended for discriminating or separating the stimulus. Amongst other signal processing methods the short time Fourier transform (STFT) for time frequency analysis and power spectrum estimation using parametric autoregressive models have also been explored in Tian *et al.* [15]. Cabral *et al.* [8] used complexity measures to characterize the nonlinearity and nonstationary behaviour of plant electrical signals inside and outside a Faraday cage. The independent component analysis (ICA) has been used in Huang *et al.* [9] to separate the superimposed response generated from epidermal, mesophyll and guard cells. Detail review of other available approaches has been discussed in Chatterjee *et al.* [7], [16] along with proposing linear/nonlinear system identification and discriminant analysis classification approaches using statistical features of segmented plant signals.

## 2. Material and Methods
### *2.1. Experimental data*

The experiments to record the plant electrical signals were conducted in a dark room to avoid any light interference. The whole setup was then placed inside a Faraday cage to limit the effect of electromagnetic interference as shown in Figure 2. The electrical signal recording was done at a rate of 10 Hz, from three points – top and middle of plant stem with the reference electrode at the base. Response of different stimulus *viz.* Sulphuric acid ($H_2SO_4$ of 5 ml, 0.05 M (mols/lit)), Sodium Chloride (NaCl of 5 or 10 ml, 3mM), and Ozone ($O_3$) was monitored using stainless steel needle (EMG like) electrodes which are then connected to a similar Data Acquisition (DAQ) system as reported in [7]. Examples of the plant electrical response due to each individual stimulus – $O_3$, $H_2SO_4$, NaCl of 5 or 10 ml are shown in Figure 1, in two different channels of a plant.

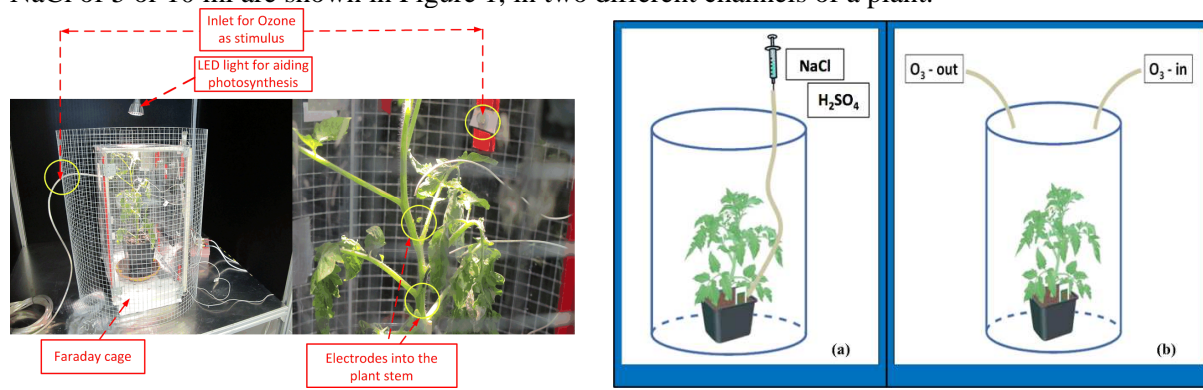

Figure 2: Experimental setup: (left) a tomato plant inside a plastic transparent box, kept inside a Faraday cage, (right) Injecting NaCl, $H_2SO_4$, $O_3$ in plants.

In the present study, we have used 11 Tomato plants (*Solanum lycopersicum*) aged between 3-5 weeks to study the effect of three different chemical stimuli - $O_3$, $H_2SO_4$, NaCl. It is understandable that there might be some possible variation due to the plant species and plant growing conditions during the recording sessions. Therefore, we have chosen plants for the experiments belonging to the same species, and they were grown in the greenhouse until the recording sessions (up to 5 weeks). For the recording sessions, each plant (at least 12 hours before the beginning of the experiment) was moved from the greenhouse to the setup within a Faraday cage that was positioned inside a box where light (day/night cycle of 12+12 hours), temperature (24°C) and humidity (60%) were controlled. The genus, species and cultivar of the tomato plants used are *Solanum lycopersicum cv. Shiren*. We used home grown tomato plants at the University of Florence, Florence, Italy. Plants grown at different environment and geographical locations may have different electrical response characteristics which is not considered in the present paper and may be explored in a future study.

### *2.2. Need of optimum digital filter for plant signal processing*

Choice of the filter is crucial in this case, because no prior knowledge of the frequency spectrum for plant's true electrical response exists. Traditionally, band-pass filters are used for most of the biological signals like ECG, EEG, EMG [1] to eliminate the effect of low frequency drift/artifact and high frequency measurement noise. Due to the 10 Hz sampling rate in our experiments, according to the Nyquist criterion, there will be no frequency component above 5 Hz which is relatively low frequency for biological signal processing [1]. Therefore, instead of removing the spectrum from both sides, we have chosen to implement a high pass filter since the higher side of the spectrum (5 Hz) is almost insensitive to measurement noise due to the low sampling rate. Whereas choosing the cut-off frequency of the high-pass filter at the lower frequency end has got more impact on shaping the frequency spectrum as well as the time domain response of the random part of the plant signals as shown in Figure 3.

This is especially important since a lower value of the filter cut-off frequency ($\omega_c$) may allow significant amount of artifact/drift to go into the plant signal. On contrary, a higher value of $\omega_c$ may remove some significant information from the frequency spectrum. In fact the main challenge lies in balancing the trade-off between loosing significant information in the frequency spectrum but not allowing low frequency drifts and artefacts to contaminate the spectrum. Figure 3 shows the effect of applying a digital high-pass Butterworth filter, as an example, with a cut-off frequency of $\omega_c = 1$ Hz for two plants with two channels. Due to the low frequency drifts present in the time domain, it is evident in Figure 3 that the frequency spectrum of the raw signals have got high power in low frequency which could mask the underlying biological response of the plant lying in the relatively higher frequencies. It is evident that the IIR filter enforces the signal to have almost zero mean and uniform variance in both the pre and post-stimulus part and therefore may help in characterizing the stimulus in terms of other higher order statistical features, similar to the exploration reported in [16]. It is to be noted that although apparently the time domain representation looks similar for the filtered signals, their frequency responses are different. Especially the gain and ripple at the low frequency region, cut-off frequency as well as type of the digital IIR filter needs to be optimally tuned, using some criterion which ensures similar statistical behaviour of the pre-stimulus part.

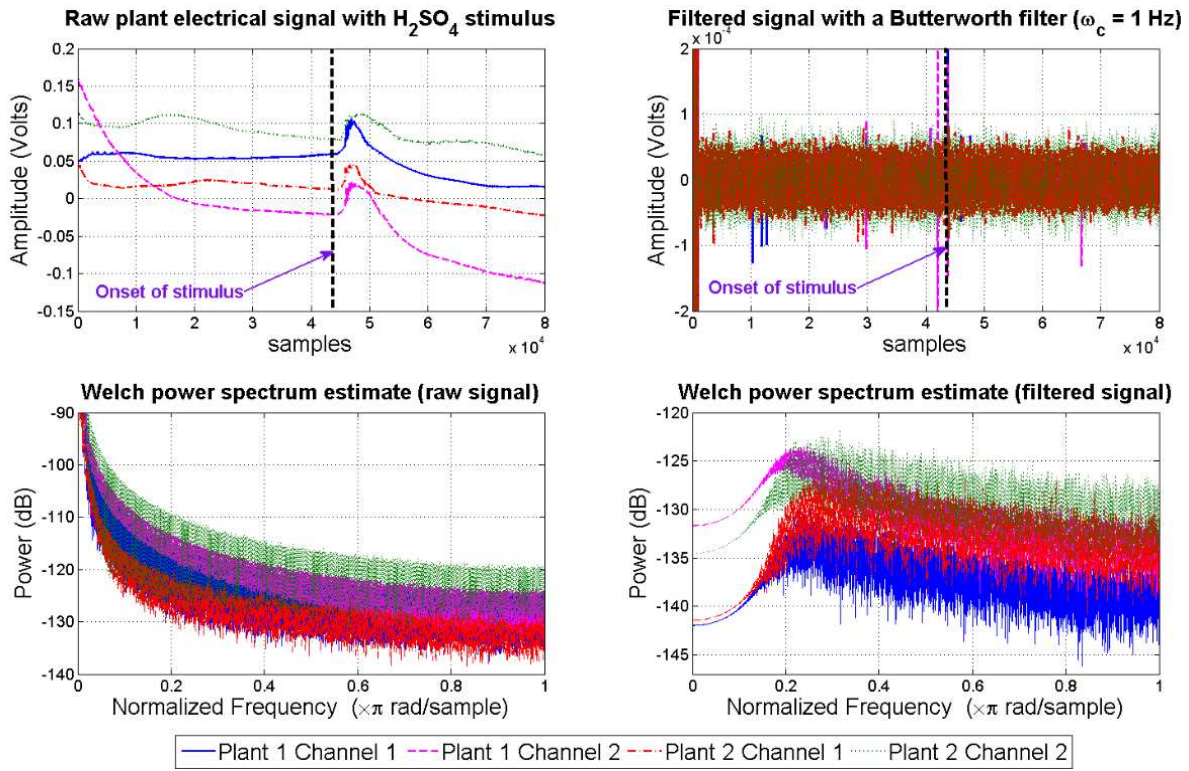

Figure 3: Time and frequency domain representation of the raw and filtered plant electrical response using a Butterworth filter with $\omega_c = 1$ Hz.

Conventionally the IIR filters are designed using the specification of the ripples of the pass and/or stop band along with the respective cut-off frequencies ($\omega_c$) to design an analog low-pass filter first. User specified values of these design parameters are traditionally used to estimate the right order of the filter. Using the design specifications, a low pass filter is designed then, followed by a transformation to map the analog low pass IIR filter to a high pass filter which is then discretized using bilinear transformation or other similar methods. In analog domain, the Butterworth filter has got smooth monotonic frequency response in both the pass and stop band. The Chebyshev type-I filter is maximally flat (smooth and monotonic) in stop band whereas the Chebyshev type-II filter is maximally flat in the pass band. Therefore, the Chebyshev type I and II filters have got equi-ripple magnitude response in the pass-band and stop-band respectively. The elliptic filter has got equi-ripple magnitude response in both the pass- and stop-band. Here, the filter specifications like cut-off frequency ($\omega_c$), optimum order ($N$), pass and stop band ripples in dB ($R_p$, $R_s$) needs to be optimized using a criterion which ensures that the statistical characteristics of the unstimulated parts of the plant signal are similar. Also a zero-phase filtering is adopted using a causal and stable filter in both forward and backward direction to get the magnitude squared response in (1).

$$\left|H_c(j\Omega)\right|^2 = H_c(s)H_c(-s)\big|_{s=j\Omega} \quad (1)$$

We now explore, four classes of analog IIR filters having magnitude squared responses given in (2).

$$\begin{aligned}
&\left|H_{Butterworth}(j\Omega)\right|^2 = 1\big/\left(1+(\Omega/\Omega_c)^{2N}\right), \\
&\left|H_{Chebyshev1}(j\Omega)\right|^2 = 1\big/\left(1+\varepsilon^2 T_N^2(x)\right), x=(\Omega/\Omega_c), \\
&T_N(x) = \begin{cases} \cos(N\cos^{-1}(x)), & |x| \leq 1 \\ \cosh(N\cosh^{-1}(x)), & |x| > 1 \end{cases} \\
&\left|H_{Chebyshev2}(j\Omega)\right|^2 = \left(\varepsilon^2 T_N^2(1/x)\right)\big/\left(1+\varepsilon^2 T_N^2(1/x)\right), \\
&\left|H_{Elliptic}(j\Omega)\right|^2 = 1\big/\left(1+\varepsilon^2 R_N^2(\Omega/\Omega_c)\right).
\end{aligned} \quad (2)$$

Here, $T_N(x)$ is the $N^{th}$ order Chebyshev polynomial, $\varepsilon$ is the ripple, $R_N(x)$ is the $N^{th}$ order Elliptic function [17]. The analog low-pass IIR filters in (2) are now transformed to form a high-pass IIR filter using the following transformation and then discretized using the bilinear transformation as given in (3), where $T_s$ is the sampling time.

$$s \rightarrow (\Omega_c/s), s = (2/T_s)\left((1-z^{-1})/1+z^{-1}\right) \quad (3)$$

### *2.3. Wavelet energy criterion for optimizing the IIR filters*

Now, the pre-stimulus parts of the plant electrical responses are decomposed using the wavelet packet transform (WPT) to represent the signals in orthonormal basis vectors to study the distribution of the energy contents in each decomposition level. The WPT, due to having its good time and frequency domain localization property, is widely used in the analysis of non-stationary biological signals [18]. A wavelet family $\psi_{a,b}(t)$ could be generated from a chosen mother wavelet $\psi(t)$ by selecting the dilation (*a*) and translation (*b*) parameters [19] as in (4).

$$\psi_{a,b}(t) = \left(1/\sqrt{|a|}\right)\psi\left((t-b)/a\right), \{a,b\} \in \Re, a \neq 0. \quad (4)$$

For our initial exploration, we choose the Daubechies (*db3*) wavelet and later also explored the robustness of the method with the selection of other mother wavelets. The IIR filtered plant signals are decomposed *via* WPT as shown in Figure 4 where the spectrum of the signal is halved at each level using low-pass and high-pass filter banks. Now, each of the wavelet packet nodes will consist equal but different parts of the spectrum and the energy content at each node can be estimated using (5), where *P* is the number of samples at each leaf node (*i*).

$$E_i = \sum_{j=1}^{P} |W_{ij}|^2, \quad i=1,2,\cdots,Q \quad (5)$$

We restrict wavelet decomposition of the IIR filtered signal up to level 2, to keep the number of the basis vectors small ($Q=4$ in this case), to make the analysis consistent and computationally efficient. After applying a chosen IIR filter (which needs further optimization), the plant signal has been segmented in smaller non-overlapping windows of $M = 256$ samples. The wavelet energy in all the four nodes at level 2 (Figure 4) for segmented signals of length *M* have been projected in a 4-dimensional feature space. We then optimize some/all of the filter parameters ($N$, $\omega_c$, $R_p$ and/or $R_s$) for the four IIR filter variants such that it enforces the centroids (in the feature space) of the pre-stimulus signals in different experiments (*D*), lie as close as possible. The objective function (*J*) is framed as the sum of the Euclidean distances of the centroids ($C_i^d$) under different experiments from the mean of all these centroids ($\mu_{C_i}$), as shown in (6).

$$J = \sqrt{\sum_{i=1}^{Q}\sum_{d=1}^{D}\left(C_i^d - \mu_{C_i}\right)^2}, \mu_{C_i} = (1/D)\sum_{d=1}^{D}C_i^d \ \forall i = 1,\cdots,Q,$$

$$C_i^d = (1/M)\sum_{m=1}^{M}E_{im} \ \forall d = 1,\cdots,D. \tag{6}$$

The objective function (6) is minimized using the Nelder-Mead Simplex algorithm with an initial guess of $\omega_c^0 = 1\,\text{Hz}, N^0 = 7, R_p^0 = 0.5\,\text{dB}, R_s^0 = 80\,\text{dB}$.

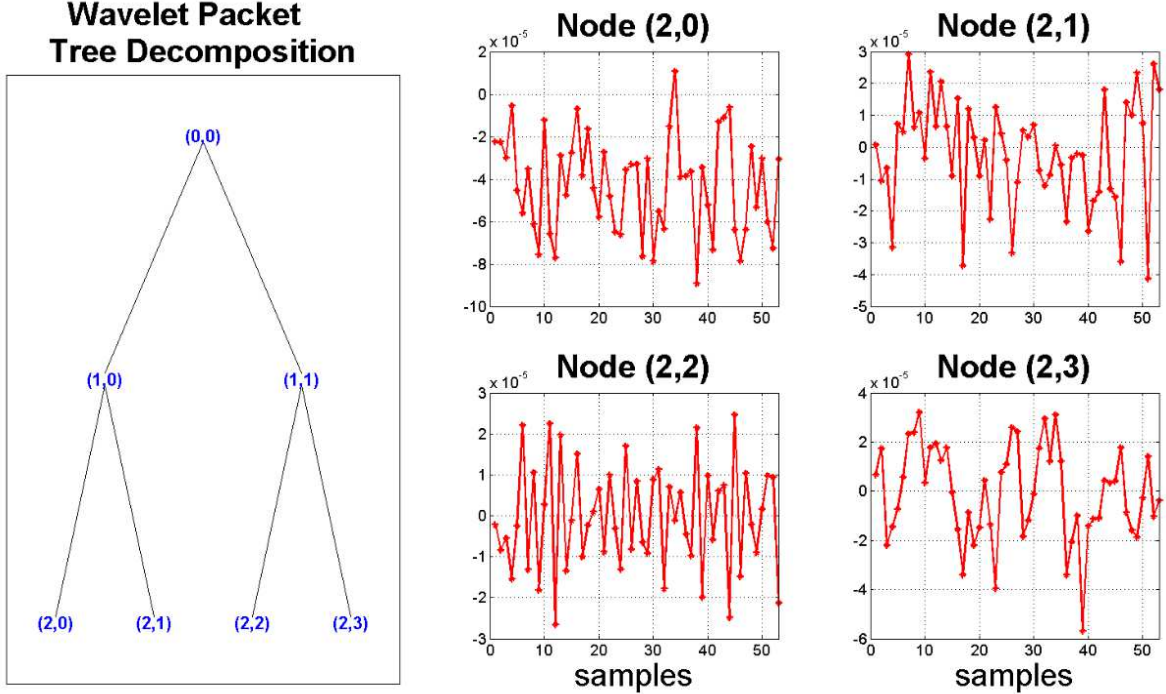

Figure 4: WPT decomposition (*db3* basis) and the coefficients in different nodes for a plant signal after Butterworth high-pass filtering with $\omega_c = 1$ Hz.

### 3. Results and Discussions
#### 3.1. Optimum IIR filter design

The optimized parameters of the IIR filters are shown in Table I for the four different structures and three data segmentation size ($M$ = 256, 512 and 1024). It is found that the Chebyshev type-II filter yields the minimum cost function ($J_{min}$) for the non-overlapping 256 samples of data segmentation. Also for the other two cases of 512 and 1024 samples data segmentation, the Chebyshev type-II filter outperforms the other three filter structures in minimizing the discrimination of the background (pre-stimulus) information, under different experimental conditions. Figure 5 confirms that for all the optimum IIR filtered pre-stimulus signals in different experimental conditions, the associated wavelet packet energy distributions (in the scatter diagram of energy at different nodes) are almost overlapped with a 256 sample data segmentation. The closeness of the centroids in the 4D feature space is also quantified as optimized cost function ($J_{min}$) in Table I.

Now once we matched the background or the pre-stimulus response of the plant using optimization, it is interesting to see whether the distribution of energy in wavelet packet nodes gets changed by the application of the external stimuli - $O_3$, $H_2SO_4$ and NaCl (5 and 10 ml both) as depicted in Figure 6. Under the same data segmentation of 256 samples and the same setting of the Chebyshev Type-II filter, it has been found that the *variance* of the wavelet energies and the shape of the distribution for different post-stimulus part*s* are *not similar*, unlike the pre-stimulus parts shown in Figure 5.

### *3.2. Test of robustness for different data segmentation size*

Investigation of different data segmentation size is also important [6, 1] in the present scenario, since under small segmentation size even non-stationary signals may exhibit stationary behaviour. The scatter diagrams of the wavelet energy (with *db3* basis) for 512 and 1024 sample data segmentation in Figure 7 also shows that the pre-stimulus parts are almost overlapped in these two cases. This shows that the optimum filter settings in Table I serves similar purpose of minimizing the discriminating ability of the background even if the data is segmented with a larger window size of 512 and 1024 samples.

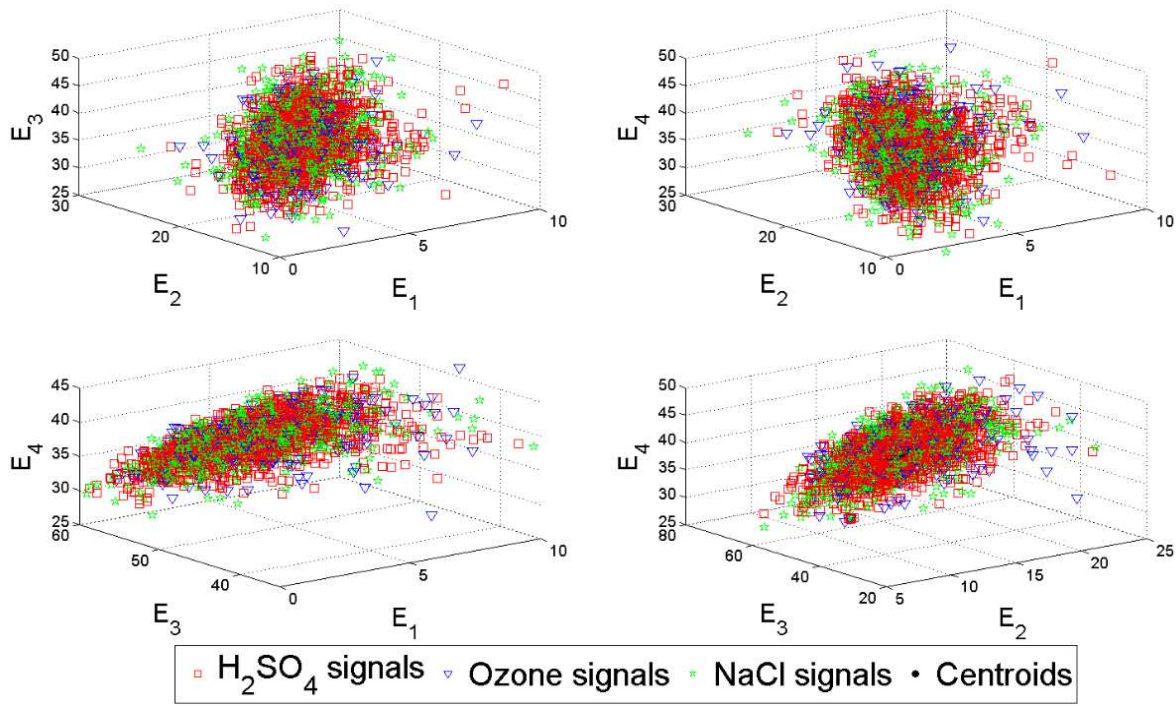

Figure 5: Overlapped wavelet energy scatterplot for the pre-stimulus signal with 256 samples data segmentation.

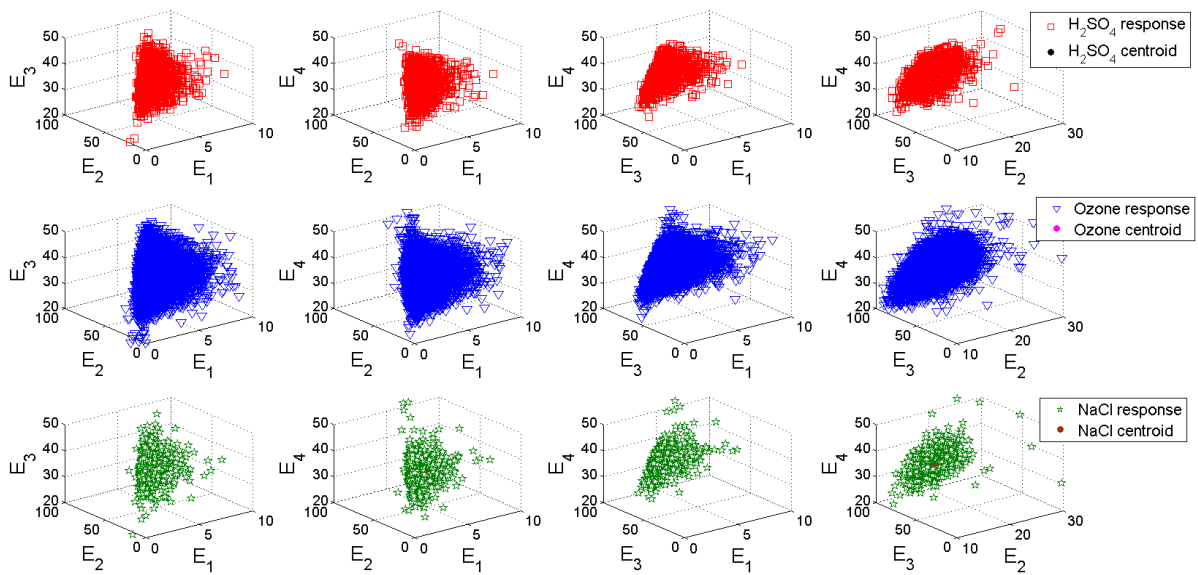

Figure 6: Increase in the spread of the post-stimulus wavelet energy with 256 samples data segmentation.

### *3.3. Test of robustness for changing the wavelet basis*

In Figure 8, box-plots of the distances amongst the centroids or the objective function in (6) have been shown, when the basis functions are changed within a chosen wavelet family [13] e.g. Daubechies (*db*), Symlet (*sym*), Coiflet (*coif*), Biospline (*bior*), and Reversebior (*rbio*). It is evident from Figure 8 that the inter-quartile range (IQR) is smallest for the *coif* family but the median is slightly higher than the *db* and *sym* family. Also, the application of *bior* and *rbio* family has led to an increase in the median and IQR respectively, and are therefore not recommended for the present application of optimum filter design for plant electrical signal processing. In addition Figure 8 shows that the relative characteristics of each wavelet family quantified as the objective function (6) remains similar but the median slightly shifts to a higher value as the data segmentation size (for feature extraction) increases from 256 to 512 and then 1024 samples.

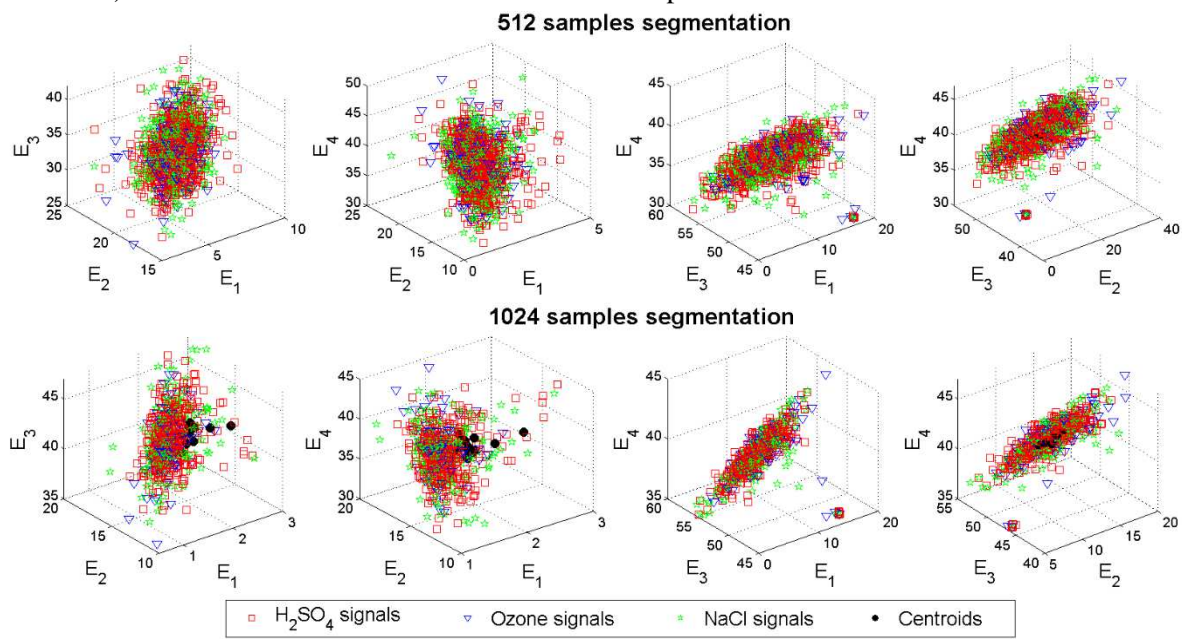

Figure 7: Effect of segmentation size (512 and 1024 samples) on the prestimulus data with *db3* basis.

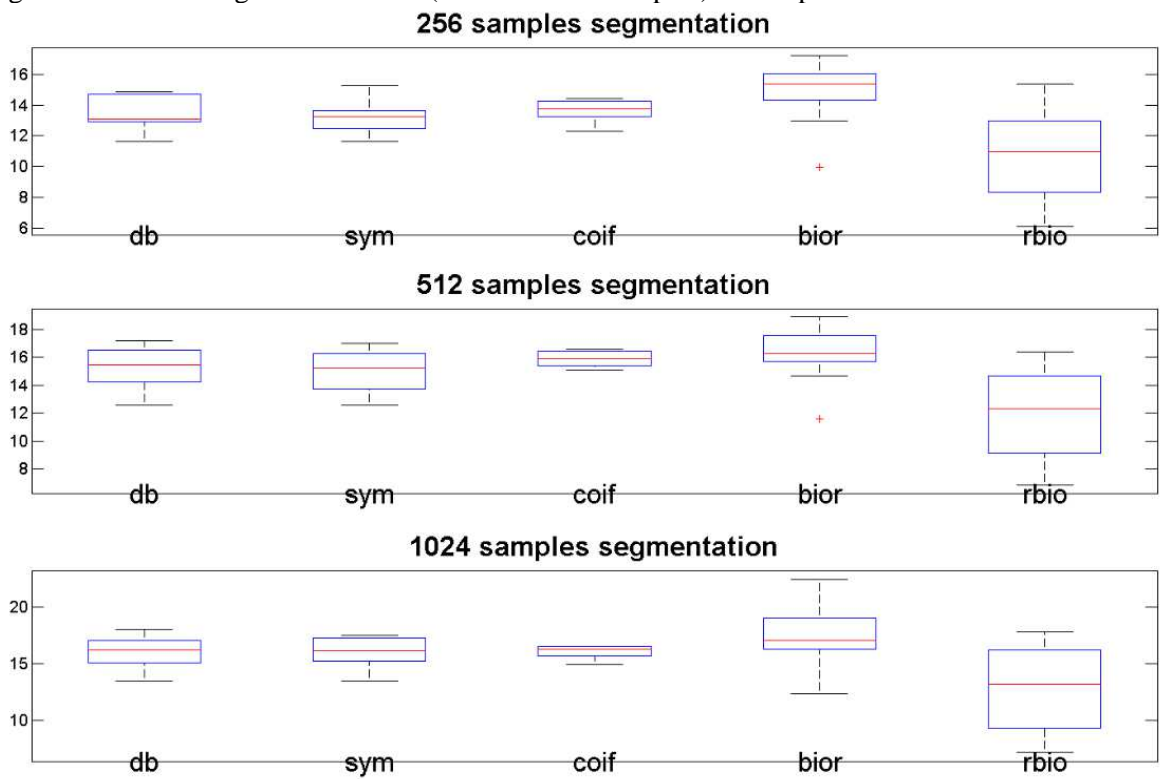

Figure 8: Box-plots of the objective function (6) for different wavelet families and data segmentation size (256, 512, amnd 1024).

As discussed, Figure 8 shows the movement of the centroids (6) within different wavelet family. Now, in order to highlight the variability in the chosen feature space (wavelet packet energy), we also report the distributions of wavlet energy in different backgrounds (pres-stimulus part in different experiments) in Figure 9. It is evident that even though the wavlet basis has changed from *db3* to *sym3*, *coif3*, *bior3.1* and *rbio3.1* in Figure 9, the data-points associated with the three different pre-stimulus signals (associated with experiments with three chemical stimuli - $O_3$, $H_2SO_4$, $NaCl$) are again almost overlapped which shows the robustness of the Chebyshev type II filter with the optimized parameters reported in Table I.

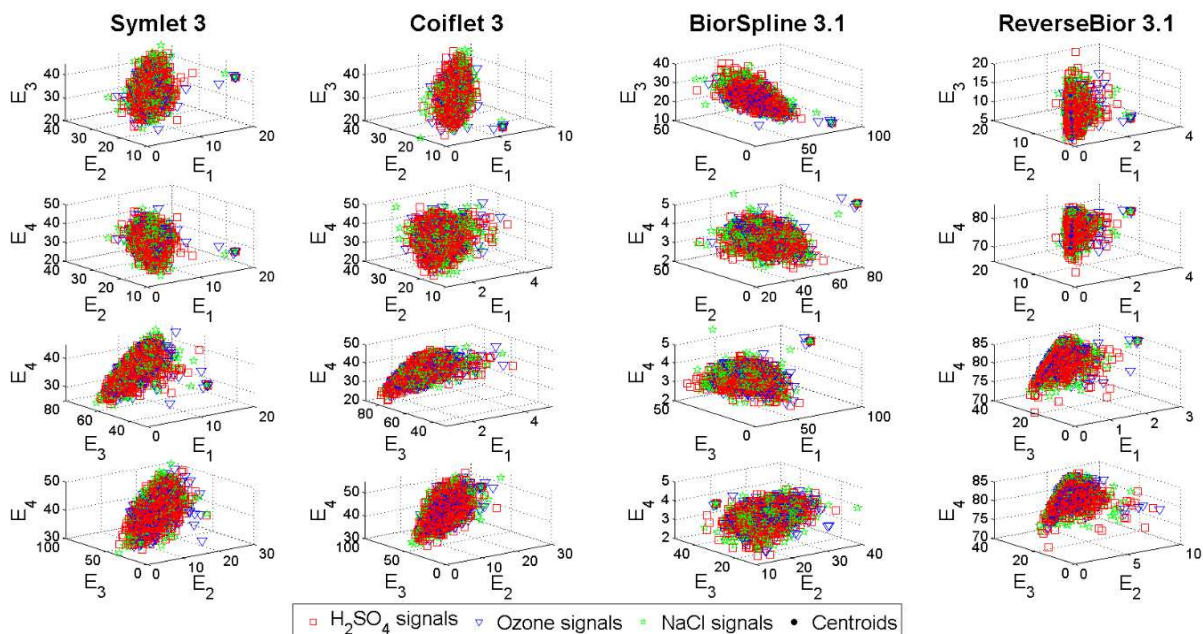

Figure 9: Backgrounds in the feature space for different stimuli with four different wavelet basis.

TABLE I
OPTIMUM IIR FILTER SETTINGS FOR DIFFERENT DATA SEGMENTATION SIZE

| Filter Type | Segment Size | $J_{min}$ | $\omega_c$ (Hz) | $N$ | $R_p$ (dB) | $R_s$ (dB) |
|---|---|---|---|---|---|---|
| Butterworth | 256 | 17.48 | 1.50 | 4 | - | - |
| | 512 | 19.77 | 1.50 | 4 | - | - |
| | 1024 | 24.41 | 1.50 | 5 | - | - |
| Chebyshev Type I | 256 | 16.71 | 1.43 | 3 | 1.00 | - |
| | 512 | 19.73 | 1.37 | 6 | 0.96 | - |
| | 1024 | 23.62 | 1.50 | 4 | 1.00 | - |
| Chebyshev Type II | *256* | *11.64* | *0.77* | *6* | - | *100* |
| | 512 | 12.55 | 0.77 | 6 | - | 100 |
| | 1024 | 13.50 | 1.34 | 6 | - | 70.19 |
| Elliptic | 256 | 17.58 | 1.45 | 6 | 0.43 | 60 |
| | 512 | 18.61 | 1.37 | 4 | 1.00 | 60 |
| | 1024 | 23.57 | 1.50 | 4 | 1.00 | 60 |

## 4. Conclusions

In this paper, we report a methodology for tuning optimum IIR filter parameters to separate out the low frequency drifts or trends, as a preprocessing step, for plant electrical signal processing applications. The effects of different data segmentation size for feature extraction and changing the wavelet basis have also been studied both graphically and quantified as the change in the proposed cost function. Future work may include characterization and classification of the externally applied chemical stimulus from the filtered or detrended plant electrical signals, instead of using the raw plant signals [16].

**Acknowledgement**

This work was supported by EU FP7 project PLants Employed As SEnsor Devices (PLEASED), EC grant agreement number 296582. The experimental data is available at http://pleased-fp7.eu/?page_id=253.